# Properties of Odd-Gap Superconductors

Elihu Abrahams

*Serin Physics Laboratory, Rutgers University, P.O. Box 849, Piscataway, New Jersey 08855*

Alexander Balatsky

*Theory Division, T-11, Los Alamos National Laboratory, Los Alamos, New Mexico 87545*
*and Landau Institute for Theoretical Physics, Moscow, Russia*

D.J. Scalapino

*Department of Physics, University of California, Santa Barbara, California 93106*

J.R. Schrieffer

*Department of Physics, Florida State University, Tallahassee, Florida 32310*

We discuss the class of superconductors which have pairing correlations which are odd in frequency, as introduced originally by Berezinskii and more recently by Balatsky and Abrahams. As follows from the equations of motion, a natural definition of the thermodynamic order parameter of the odd-pairing state is the expectation value of a composite operator which couples a Cooper pair to a spin or charge fluctuation. We use a model pairing hamiltonian to describe properties of the odd-pairing composite-operator condensate. We show that the superfluid stiffness is positive, we discuss superconductive tunneling with an ordinary superconductor and we derive other thermodynamic and transport properties.

## I. INTRODUCTION

Novel symmetry types of the energy gap function $\Delta(\mathbf{k}, \omega)$ in spin-singlet superconductors were discussed recently[1] in the context of high-$T_c$ superconductivity,[2,3] in order to investigate ways of constructing a superconducting state which is not inhibited by the the presence of strong short-range repulsion. An anisotropic superconducting gap parameter ("$p$-wave" or "$d$-wave") which changes sign as $\mathbf{k}$ moves around the Fermi surface is the simplest way of achieving such a state. An alternative posssibility was proposed some time ago by Berezinskii,[4] who pointed out in the context of $^3$He that it is permissable to have triplet ($S = 1$) pairing in which the gap function $\Delta(\mathbf{k}, \omega)$ is even in momentum $\mathbf{k}$ (for example "$s$-wave"), provided it is odd in frequency $\omega$. Recently, this suggestion was extended to the singlet case,[1] where it was argued that a new class of singlet superconductors can exist for which the gap function is an odd function of both $\mathbf{k}$ and $\omega$. Such a superconducting state is thus odd under both parity and time reversal and, as in the case discussed in Refs. 2, 3, the effect of short-range repulsion is suppressed, here because the equal-time pair amplitude is zero.

An odd-in-frequency solution to the Eliashberg equations can only exist for certain forms of the pairing kernel which scatters a pair from $\mathbf{k}, -\mathbf{k}; \omega, -\omega$ to $\mathbf{k}', -\mathbf{k}'; \omega', -\omega'$. In the singlet case, to obtain an odd $\omega$ gap parameter, the kernel must have a contribution which is odd in both $\mathbf{k}$ and $\omega$. The problem of how an appropriate interaction might arise has been discussed in a recent paper[5] where a specific example based on spin-fluctuation exchange in the two-dimensional Hubbard model was discussed. If an interaction admits an odd solution it still must be checked whether it only describes a metastable phase with free energy higher than that of an even-frequency solution.

In the present paper, we explore another issue which arises because there is an instability which was not mentioned[6] in all the known translationally invariant odd-gap solutions considered in Refs. (1,4). Namely, because of the symmetry properties of the odd-gap pairing amplitudes, there is a change in sign of the anomalous parts of BCS coherence factors. A consequence is that a calculation of the Meissner effect in lowest order[7] gives an opposite sign to that of the usual BCS case so that the superfluid density appears to be negative. Thus, although the free energy of the uniform odd gap state is less than that of the normal state,[1] the system prefers a non-uniform distribution of supercurrents, i.e. some sort of vortex state which could in principle be determined in a calculation which goes beyond the linearized Eliashberg equation which is used to find $T_c$.

In this paper, we shall show how a stable Meissner state may be achieved by the introduction of a composite condensate which is characterized by the joint condensation of a Cooper pair and a density fluctuation. Our point of view is motivated by the question: If the anomalous Green's function vanishes at equal times, as it does for an odd gap, what is the appropriate macroscopic thermodynamic order parameter? For example, in the usual BCS case it is the expectation value $F(\mathbf{r}, t; \mathbf{r}', t' \to t) = \langle \psi(\mathbf{r}, t)\psi(\mathbf{r}', t)\rangle$ for $|\mathbf{r} - \mathbf{r}'| < \xi$, where $\xi$ is the BCS coherence length. Since this vanishes in the odd case (it is odd in $t - t'$), we are led naturally to consider the quantity $[dF(\mathbf{r}, t; \mathbf{r}', t')/dt]$ for $t \to t'$ as the equal-time correlation function to be used as order parameter. Given a pairing hamiltonian, this quantity can be obtained from the equations of motion and it consists of the expectation value of the product of a pair operator and, depending on the nature of the pairing interaction, a spin or charge density fluctuation. This quantity is of the form

$$\langle M(\mathbf{r}, t) \cdot \psi(\mathbf{r}, t)\psi(\mathbf{r}, t' \to t)\rangle, \tag{1.1}$$

with an appropriate spin structure which will be discussed below. The operator $M$ contains a spin or charge density fluctuation:

$$M(\mathbf{r}, t) = \int d\tau d\mathbf{r}' \, \psi^\dagger(\mathbf{r}', t + \tau)\psi(\mathbf{r}', t + \tau) \, \mathcal{D}(\mathbf{r} - \mathbf{r}', \tau), \tag{1.2}$$

where $\mathcal{D}$ is a generalized non-local retarded interaction.

In Section II, we present a symmetry analysis of the Green's functions which leads to this new type of superconductor. We show that for the spin-singlet (triplet) case, the only symmetry requirement is that the gap function $\Delta(\mathbf{k}, \omega)$ be even (odd) under simultaneous change of sign of momentum and frequency. Consequently, apart from the conventional even-frequency gaps which are even (odd) in momentum for singlet (triplet) pairing, there can be solutions of the gap equation which are *odd (even)* in momentum and odd in frequency in the singlet (triplet) case.

In Section III, we give an illustration of how a composite operator arises naturally as the order parameter for an example system. For purposes of calculation, we introduce a model hamiltonian which can lead to odd-frequency pairing and we derive the gap equation near the critical temperature in terms of the parameters of the model. We find that the strength of the interaction must exceed a critical value in order that the odd solution exist. This is a rather general property of odd-gap solutions[4,3] and is in contrast to the BCS case where the logarithmic behavior of the gap equation insures a non-trivial solution no matter how weak the coupling. In our case, the kernel of the gap equation, being odd in $\omega$, must vanish at zero frequency if the kernel varies smoothly in this vicinity; this removes the logarithmic divergence.

In Section IV, we calculate the Green's functions in the composite operator condensate and show explicitly that the superfluid stiffness is positive. Section V contains a discussion of the Josephson effect between odd-gap and even-gap superconductors. In Section VI we calculate the NMR relaxation rate and in Section VII the effect of impurity scattering for the odd-gap composite superconductor

Subsequent to the singlet odd-gap discussion of Balatsky and Abrahams,[1] several authors, in particular Kivelson and Emery,[8] Balatsky and Bonča,[9] and Coleman, Miranda and Tsvelik,[10] have proposed other scenarios for odd pairing involving composite operators. These examples indicate that the odd-frequency-gap superconducting correlations are relevant for some physical models. Although a robust realization of odd-gap superconductivity remains to be found, in what follows we consider a simple model in which some properties of the odd gap superconductors, such as the Meissner effect, NMR and Josephson effects, can be analyzed.

## II. SYMMETRY OF THE GAP

We introduce the anomalous Green's function

$$\mathcal{F}(\mathbf{k}, \omega_n) = \sum_{\alpha,\beta} \int_0^{1/T} d\tau \, e^{-i\omega_n \tau} \, \langle T_\tau c_{\mathbf{k},\alpha}(\tau/2) c_{-\mathbf{k},\beta}(-\tau/2)\rangle \cdot \phi_{\alpha\beta}(\mathbf{k}), \tag{2.1}$$

where $\omega_n = (2n + 1)\pi T$ and $\phi_{\alpha\beta}(\mathbf{k})$ determines the spin symmetry of the pairing amplitude and hence the gap function. For example, for singlet pairing, $\phi_{\alpha\beta} = i(\sigma^y)_{\alpha\beta}/2$, where $\sigma^y$ is the Pauli spin matrix. For triplet pairing (as in $^3$He, for example) $\phi_{\alpha\beta}(\mathbf{k}) = (i\sigma^y \vec{\sigma})_{\alpha\beta} \cdot \mathbf{d}(\mathbf{k})$, where $\mathbf{d}(\mathbf{k})$, a vector in spin space, is a linear combination of the three $Y_{\ell m}(\hat{\mathbf{k}})$ for $\ell = 1$, each of which corresponds to one of the $m_z = \pm 1, 0$ states. For example, the form $\hat{\mathbf{d}}_x \pm i\hat{\mathbf{d}}_y$ gives the $m = \pm 1$ component of the triplet. Similar forms hold for the anomalous self energy $W(\mathbf{k}, \omega_n)$ and the gap function $\Delta(\mathbf{k}, \omega_n)$.[11]

The *only* constraint on the possible symmetry of $\mathcal{F}(\mathbf{k}, \omega_n)$ and $W(\mathbf{k}, \omega_n)$ follows from the anticommutativity of the $\psi$ operators in $\mathcal{F}$. Using this and the antisymmetry of the Pauli matrices we immediately get, for the singlet case:[11]

$$\mathcal{F}(\mathbf{k}, \omega_n) = \mathcal{F}(-\mathbf{k}, -\omega_n), \quad \Delta(\mathbf{k}, \omega_n) = \Delta(-\mathbf{k}, -\omega_n) \tag{2.2}$$

and for the triplet[4]

$$\mathcal{F}(\mathbf{k},\omega_n) = -\mathcal{F}(-\mathbf{k},-\omega_n), \quad \Delta(\mathbf{k},\omega_n) = -\Delta(-\mathbf{k},-\omega_n) \tag{2.3}$$

From these relations, it can be seen that the conventional even spatial parity singlet (conventional metals) and odd parity triplet ($^3$He) states have gap functions which are *even* in Matsubara frequency. Then the equal time anomalous Green's function $\mathcal{F}$ is nonzero and this leads to the usual off-diagonal long-range order, ODLRO. However it has been shown[1,4,5] that gap functions which are *odd* in Matsubara frequency are admissible. In the odd case, Eq. (2.2) is satisfied with singlet superconducting pairing with a pair amplitude which is odd in both $\mathbf{k}$ and $\omega_n$:

$$\begin{aligned}\mathcal{F}(\mathbf{k},\omega_n) &= -\mathcal{F}(-\mathbf{k},\omega_n) = -\mathcal{F}(\mathbf{k},-\omega_n) \\ \Delta(\mathbf{k},\omega_n) &= -\Delta(-\mathbf{k},\omega_n) = -\Delta(\mathbf{k},-\omega_n).\end{aligned} \tag{2.4}$$

Similarly, for the triplet case, Eq. (2.3) is satisfied for triplet superconducting pairing with a gap which is even in $\mathbf{k}$ and odd in $\omega_n$:

$$\begin{aligned}\mathcal{F}(\mathbf{k},\omega_n) &= \mathcal{F}(-\mathbf{k},\omega_n) = -\mathcal{F}(\mathbf{k},-\omega_n) \\ \Delta(\mathbf{k},\omega_n) &= \Delta(-\mathbf{k},\omega_n) = -\Delta(\mathbf{k},-\omega_n).\end{aligned} \tag{2.5}$$

Eqs. (2.4, 2.5) show that the spin-singlet (triplet) gap is described in terms of an odd (even) orbital function, while, at the same time, the spin function is odd (even). There is no violation of the Pauli principle because the equal-time gap function, or pair amplitude, for singlet or triplet vanishes since the gap is odd in $\omega_n$.[12]

## III. EQUATIONS FOR COMPOSITE ODD SUPERCONDUCTIVITY

### A. Model for Composite Operator

For simplicity, we consider the $m = 1$ triplet Berezinskii[4] state. The conclusions are similar for the singlet case, but the notation is somewhat more cumbersome. From Eq. (2.1), we write the anomalous Green's function as

$$\mathcal{F}(\mathbf{k},\tau) = \sum_{\alpha\beta}\langle T_\tau c_{\mathbf{k},\alpha}(\tau)\, c_{-\mathbf{k},\beta}(0)\rangle (i\vec{\sigma}\sigma^y)_{\alpha\beta}\cdot\mathbf{d}). \tag{3.1}$$

which, for $\mathbf{d} = -(\hat{\mathbf{d}}_x + i\hat{\mathbf{d}}_y)/2$ (for example), reduces to

$$\mathcal{F}(\mathbf{k},\tau) = \langle T_\tau c_{\mathbf{k},\uparrow}(\tau)\, c_{-\mathbf{k},\uparrow}(0)\rangle. \tag{3.2}$$

We are considering the odd gap case in which $\mathcal{F}(\tau)$ is odd in $\tau$ and even in $\mathbf{k}$, an even parity triplet odd gap. Since $\mathcal{F}(\tau)$ is odd, it vanishes at $\tau = 0$. However, its $\tau$-derivative is even and is generally non-zero at $\tau = 0$.[13] Therefore, we study the anomalous amplitude

$$\dot{\mathcal{F}} = \frac{\partial\mathcal{F}(\mathbf{k},\tau)}{\partial\tau} = \langle T_\tau \frac{\partial c_{\mathbf{k}\uparrow}(\tau)}{\partial\tau}\, c_{-\mathbf{k}\uparrow}(0)\rangle. \tag{3.3}$$

The behavior of $\dot{\mathcal{F}}$ is deduced from the equations of motion. The hamiltonian is

$$\mathcal{H} = H_c + H_S + H_X, \tag{3.4}$$

where $H_c$ is the kinetic energy of the conduction electrons, $H_S$ is the hamiltonian for some spin or charge excitation and $H_X$ couples the electrons to the excitations. If one integrates out the excitations to produce an effective retarded (perhaps spin-dependent) electron-electron interaction as in the Eliashberg[14] treatment of superconductivity, then $\dot{\mathcal{F}}$ can be evaluated by commuting $c_{\mathbf{k}\uparrow}(\tau)$ with the effective Hamiltonian. In that case, $\dot{\mathcal{F}}$ has a term of the form of Eq. (1.1).

In order to motivate the structure of the odd composite condensate, we consider a spin fermion model in which the conduction electrons are coupled via $H_X$ to localised spin modes which arise from a second band of electrons. Thus, $H_X$ has the form

$$H_X = J \sum_i c_{i\alpha}^\dagger \ \vec{\sigma}_{\alpha\beta} \ c_{i\beta} \cdot \mathbf{S}_i, \tag{3.5}$$

where $\mathbf{S}_i$ is a spin 1 excitation at site $i$ of the lattice. In what follows, we shall assume that both the possibility of Kondo screening (for $J > 0$) and the induced ("RKKY") interaction between the local spins can be neglected in the present context.

As in Eq. (3.3), the pairing is studied by means of the anomalous amplitude $\dot{\mathcal{F}}_i(\tau)$ which is determined by $\dot{c}_i = [H_c + H_X, c_i]$:

$$\dot{\mathcal{F}}_i(\tau) = -\sum_j t_{ij} \langle T_\tau c_{j\uparrow}(\tau) c_{i\uparrow}(0) \rangle - J\vec{\sigma}_{\uparrow\beta} \cdot \mathbf{L}_{i\beta}(\tau), \tag{3.6}$$

where $t_{ij}$ is the lattice kinetic energy matrix element and $\mathbf{L}_{i\beta}(\tau)$ is the composite anomalous amplitude which characterizes the condensate. $\mathbf{L}$ is a vector in spin-space and for the $m = 1$ odd condensate, it has the form $\mathbf{L}_{i\beta}(\tau) = \langle T_\tau c_{i\beta}(\tau) c_{i\uparrow}(0) \mathbf{S}_i(\tau) \rangle$. We shall be interested in a spin-one condensate in which a spin-zero Cooper pair is coupled to a spin-one excitation (magnon) $\mathbf{S}^\pm$. Therefore, we shall study

$$\mathbf{L}(1,2,3) = \mathbf{L}_{\alpha\beta}(1,2,3)\varepsilon_{\alpha\beta}/\sqrt{2} = \langle T_\tau c_{1\alpha}(\tau_1) c_{2\beta}(\tau_2) \mathbf{S}_3(\tau_3) \rangle \varepsilon_{\alpha\beta}/\sqrt{2}, \tag{3.7}$$

where 1,2,3 label space and imaginary time points and $\varepsilon_{\alpha\beta}$ is $i\sigma_{\alpha\beta}^y$.

### B. Reduced Hamiltonian and Mean-Field Equations

Our aim is to calculate properties of the odd pairing state which depend on the structure and symmetry of the composite order parameter. For this purpose, we start with the simplest possible weak-coupling model which gives rise to a condensate characterized by a non-zero value of $\mathbf{L}$, the expectation value of the composite operator discussed in the previous subsection and which we can solve analytically. In direct analogy to BCS theory,[15] we write a "reduced" hamiltonian

$$H_{red} = H_0 + \sum_{ij} c_{i\alpha}^\dagger c_{i\beta}^\dagger \mathbf{S}_i \cdot V_{ij} \ c_{j\gamma} c_{j\delta} \mathbf{S}_j^\dagger \ (\varepsilon_{\alpha\beta}\varepsilon_{\delta\gamma}). \tag{3.8}$$

Here $V_{ij}$ is an attractive (i.e. $V < 0$) short-range, instantaneous interaction which mediates the condensation and $H_0$ contains the kinetic energy of quasiparticles and the non-interacting hamiltonian for the spins.

We should stress that the above hamiltonian is only an instantaneous approximation to a dynamical six-point interaction. A dynamical theory which takes into account the retardation in $\mathbf{L}(1,2,3)$ can be constructed, for example starting with the interaction $H_X$ of the previous subsection [Eq. (3.5)][16,17] and perhaps other couplings as well.

For the reasons stated above, we work with $H_{red}$. Such a reduced hamiltonian might be obtained as an approximation to the repeated interparticle scattering of two quasiparticles with a spin, i.e. the Faddeev scattering of two quasiparticles with a spin fluctuation. Alternatively, if we take the simple interaction of the previous subsection and add an electron-electron interaction $U$, then the irreducible six-point coupling of $H_{red}$ could arise as shown in the diagram of Fig. 1.

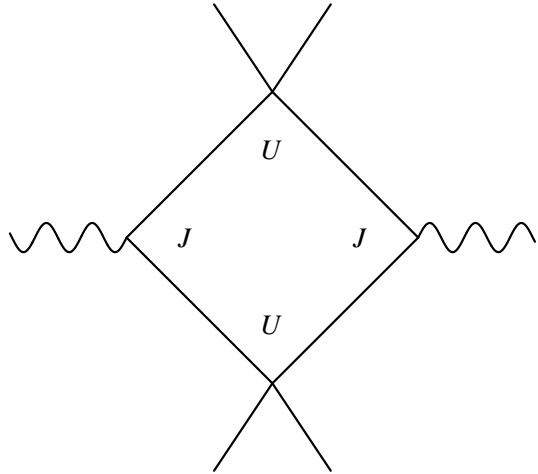

FIG. 1. Six point coupling arising from electron-electron interaction ($U$) and electron-localised spin coupling ($J$).

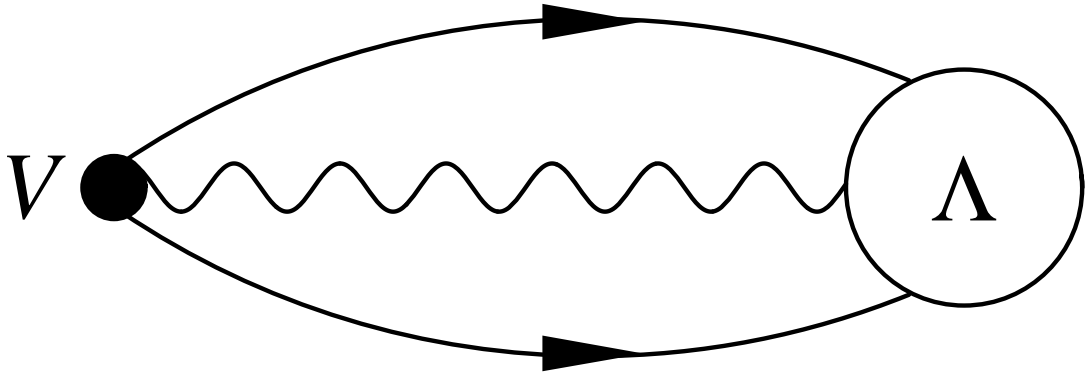

FIG. 3. Linearized gap equation for composite condensate

The gap equation differs from the BCS case by the presence of the spin propagator in the integral. It is essential to recognize that the boson described by $\mathcal{D}$ is *not* what is responsible for forming the Cooper pair part of the condensate. The interaction for the condensate here is the instantaneous $V$. We shall show in the next subsection that for a simple choice for the propagator $\mathcal{D}$, provided $V$ exceeds a minimum coupling, there is an instability which determines $T_c$.

## C. Behavior Near the Transition Temperature

In contrast to the BCS case, the mean-field hamiltonian, Eq. (3.9), is not quadratic in the fields so that an immediate diagonalization is not possible. Instead, we shall assume that the transition is second order everywhere so that we may construct a perturbation theory in $\mathbf{\Lambda}$, which should be valid both near $T_c$ and near the minimum coupling, where $\mathbf{\Lambda}$ is small. This approach is similar to that used by Gor'kov in the microscopic derivation[7] of Ginzburg-Landau equations. In this way, we can obtain the transition temperature and various response functions close to the transition. The elementary vertices $\mathbf{\Lambda}$, $\tilde{\mathbf{\Lambda}}$ are shown in Fig. 4.

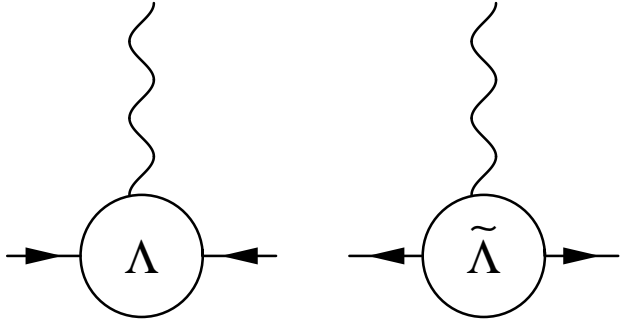

FIG. 4. Elementary anomalous vertices for perturbative treatment.

We begin with the anomalous condensate amplitude $\mathbf{L}_i(\tau)$. To third order, it is given by the diagrams in Fig. 5 and the analytic expression is

$$\mathbf{L}(1,2,3) = \int d4 \; \mathcal{G}(1,4)\mathbf{\Lambda}(4)\mathcal{G}(2,4)\mathcal{D}(3,4) \tag{3.12}$$

$$+ \int d4d5d6 \; \mathcal{G}(1,4)\mathbf{\Lambda}(4) \cdot \mathcal{G}(5,4)\tilde{\mathbf{\Lambda}}(5)\mathcal{G}(5,6)\mathbf{\Lambda}(6)\mathcal{G}(2,6)\mathcal{D}(4,6)\mathcal{D}(3,5).$$

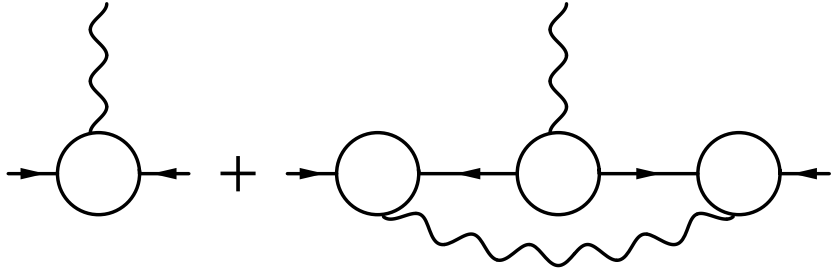

FIG. 5. Perturbation expansion for $\mathbf{L}$.

The propagators $\mathcal{G}$ and $\mathcal{D}$ are for quasiparticles and spins in the normal state. The transition temperature is found from Eq. (3.10) and the linearized equation for $\mathbf{L}$ [compare Eq. (3.11)]:

$$\mathbf{\Lambda}(1) = V\mathbf{L}(1,1,1) = V \int d2 \; \mathcal{G}(1,2)\mathbf{\Lambda}(2)\mathcal{G}(1,2)\mathcal{D}(1,2). \tag{3.13}$$

Here we have taken the interaction $V$ to be of zero range. The $T_c$ equation is therefore given by

$$1 = VT_c^2 \sum_{\mathbf{k},\mathbf{q}} \sum_{\omega,\nu} \mathcal{G}(\mathbf{k},\omega)\mathcal{G}(-\mathbf{k}+\mathbf{q},-\omega+\nu)\mathcal{D}(\mathbf{q},\nu), \tag{3.14}$$

where $\omega$, $\nu$ are fermion (odd) and boson (even) Matsubara frequencies respectively.

The precise result for $T_c$ depends on the form of the magnon propagator. It is obvious that $\mathcal{D}(1,2) = $ const leads to an ordinary BCS expression for $T_c$. To introduce the new physics of the composite condensate we choose $\mathcal{D}(\mathbf{q},\nu) = \Gamma/[(i\nu)^2 - \Gamma^2]$, which could represent the dynamics of non-interacting localized Kondo-like spin resonances. Since $\mathcal{D}$ is independent of momentum, we carry out the momentum sums in Eq. (3.14) and find

$$1 = -2\pi^2 N_0^2 VT^2 \sum_{\omega>0,\;\omega'>0} [\mathcal{D}(\omega+\omega') - \mathcal{D}(\omega-\omega')], \tag{3.15}$$

where $N_0$ is the one-spin quasiparticle density of states at the Fermi surface. We shall see below that the sums are dominated by their low-frequency behavior. Therefore, for simplicity, we approximate the square bracket of Eq. (3.15), i.e. the odd part of $\mathcal{D}$, by the separable form

$$2\mathcal{D}_{odd} \simeq 4\Gamma \frac{\omega}{\omega^2 + \Gamma^2} \cdot \frac{\omega'}{\omega'^2 + \Gamma^2}. \tag{3.16}$$

Eq. (3.15) can now be evaluated explicitly for $T_c$. In terms of digamma functions, the result is

$$1 = g\mathrm{Re}\left[\psi(\frac{1}{2} + \frac{\omega_c + i\Gamma}{2\pi T_c}) - \psi(\frac{1}{2} + \frac{i\Gamma}{2\pi T_c})\right]^2, \tag{3.17}$$

where $g = 2N_0^2 V\Gamma$ and $\omega_c$ is the ultraviolet cutoff of the interaction. In Fig. 6, we show a plot of $T_c$ against the effective coupling $g$. As expected, there is a minimum value of $g$ to achieve a transition and the curve suggests that for a range of $g$ above the minimum value, there is reentrant behavior. This may be checked by evaluating the right-hand side (RHS) of Eq. (3.14) near $T_c$. The condensed state occurs for those temperatures, larger or smaller than $T_c$ for which the temperature derivative of the RHS is negative. In this way, we find that in the (smaller coupling) region of Fig. 6 where $T_c$ has two values $T_{c+}$, $T_{c-}$ the condensate exists between them and the system is normal for $T < T_{c-}$ or $T > T_{c+}$. In the larger coupling region, only $T_{c+}$ exists and the system is in a condensed phase down to $T = 0$.

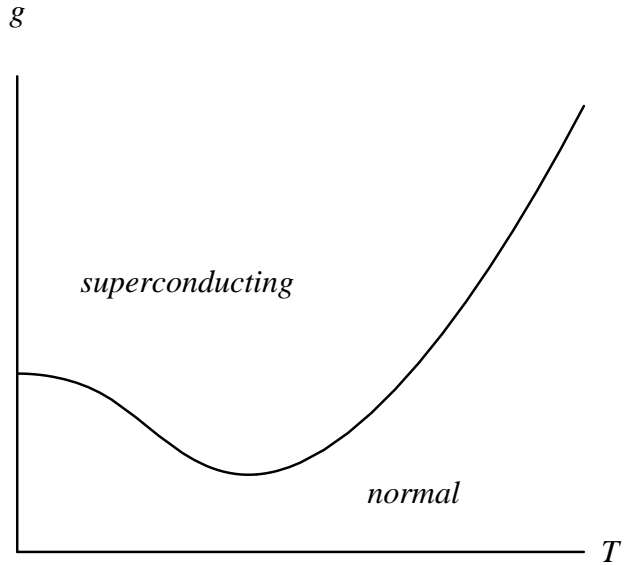

FIG. 6. Phase diagram for composite odd gap state, showing reentrant behavior.

Another possibility is a spatially uniform magnon propagator of the form

$$\mathcal{D}(\mathbf{q}, \nu) = -\delta_\mathbf{q} \frac{\Gamma}{(i\nu)^2 - \Gamma^2} \tag{3.18}$$

In this case, the momentum delta function removes one integration and a infrared logarithmic integral (as in BCS) is recovered. The $T_c$ equation becomes

$$1 = 2N_0 V T_c^2 \sum_{\omega > 0,\ \omega' > 0} \left[\frac{\Gamma}{(\omega - \omega')^2 + \Gamma^2}\frac{1}{\omega + \omega'}\right] \tag{3.19}$$

The result of the frequency sums gives the transition temperature as

$$T_c = 1.56\omega_c e^{-\pi/N_0 V}. \tag{3.20}$$

In order to find the quasiparticle spectrum in the condensed state, we examine the single-particle Green's function. For this, we calculate the irreducible self-energy to second order in $\Lambda$. The diagram is given in Fig. 7 and the expression is

$$\Sigma(1,2) = (4/3)\mathbf{\Lambda}(1) \cdot \tilde{\mathbf{\Lambda}}(2)\mathcal{G}(2,1)\mathcal{D}(1,2). \tag{3.21}$$

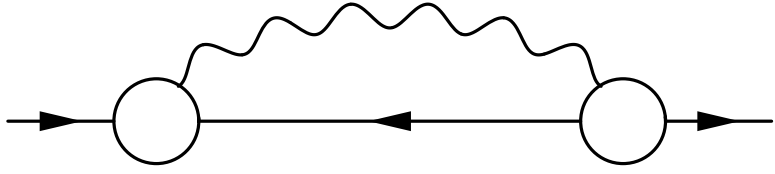

FIG. 7. Diagram for self energy.

After Fourier transform, Eq. (3.18) takes the form

$$\Sigma(\mathbf{k},\omega) = (8\pi i/3)N_0|\mathbf{\Lambda}|^2 T \sum_{\omega'>0} \mathcal{D}_{odd}(\omega+\omega') \tag{3.22}$$

With the use of the separable propagator of Eq. (3.16), we find

$$\mathcal{G} = \mathcal{G}^0[1+\Sigma\mathcal{G}] = \frac{1}{i\omega Z(i\omega) - \epsilon}, \tag{3.23}$$

where $\mathcal{G}^0$ is the normal state Green's function, $\epsilon$ is the normal state quasiparticle kinetic energy measured from the Fermi energy, and

$$Z(i\omega) = 1 - W^2/[(i\omega)^2 - \Gamma^2], \quad W^2 = (4\pi/3)|\mathbf{\Lambda}|^2\sqrt{\Gamma/2V}. \tag{3.24}$$

From Eq. (3.23), we find that the density of states in the superconducting state is unchanged from that of the normal state, a situation which always obtains when the self-energy is momentum independent (as in the case e.g. of electron-phonon interactions). However, the interaction produces mass renormalization. Here, there is no gap, but the quasiparticle spectrum is modified to the form

$$E(\mathbf{k}) \simeq \epsilon(\mathbf{k})[1 + \frac{W^2}{\epsilon(\mathbf{k})^2 - \Gamma^2}], \tag{3.25}$$

which is valid near the transition where $|\mathbf{\Lambda}|^2$ and hence $W^2$ is small. For small excitation energy, there is a dynamical mass enhancement which does not appear in the density of states, but which is important in response functions.

## IV. MEISSNER EFFECT

The Meissner effect occurs when the paramagnetic electrodynamic response is less than the diamagnetic part. The dc response is given by

$$\mathbf{j}_i(\mathbf{q}) = -Q_{ij}(\mathbf{q}) A_j(\mathbf{q})$$
$$Q_{ij}(\mathbf{q}) = \delta_{ij} Ne^2/m + Q_{ij}^p(\mathbf{q}) \tag{4.1}$$

The paramagnetic kernel is[15]

$$Q_{ij}^p(\mathbf{q}) = -(e^2/4m^2)\sum_{\alpha\beta}\sum_{\mathbf{k}\mathbf{k}'} \mathbf{k}_i \mathbf{k}'_j \int_{-\beta}^{\beta} d\tau \langle T c_\alpha^\dagger(\mathbf{k}_+,\tau)c_\alpha(\mathbf{k}_-,\tau)c_\beta^\dagger(\mathbf{k}'_-,0)c_\beta(\mathbf{k}'_+,0)\rangle. \tag{4.2}$$

where $\mathbf{k}_\pm = \mathbf{k} \pm \mathbf{q}/2$.

As in the previous section, we evaluate $Q_p$ near $T_c$ by perturbation in the order parameter $\mathbf{\Lambda}$. The relevant diagrams are shown in Fig. 8.

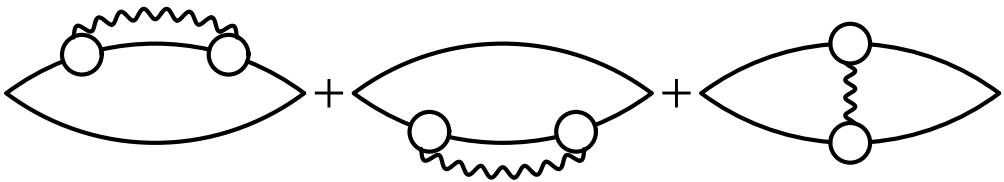

FIG. 8. Current-current correlation function for the Meissner effect.

The analytic expression for $\mathbf{q} \to 0$ is

$$Q_{ij}^p - Q_{ij}^n = \frac{e^2 T^2}{m^2}|\Lambda|^2 \sum_{\omega\omega',\mathbf{k}\mathbf{k}'} k_i k_j' \ [\mathcal{G}^2(\mathbf{k},\omega)\mathcal{G}^2(\mathbf{k}',\omega') - 2\mathcal{G}^3(\mathbf{k},\omega)\mathcal{G}(\mathbf{k}',\omega')]\mathcal{D}(\mathbf{k}+\mathbf{k}',\omega+\omega'). \tag{4.3}$$

A result $Q^p - Q^n > 0$ indicates a Meissner effect and a positive superfluid density.

If the magnon propagator is momentum independent, there is no contribution to the Meissner kernel since the momentum summands are odd. Therefore, we generalize the magnon propagator to include momentum dependence. An extreme possibility is the case of a static, spatially uniform magnon of the factorized form

$$\mathcal{D}(\mathbf{q},\nu) = -\delta_\mathbf{q} \ \delta_\nu/T. \tag{4.4}$$

When this is used in Eq. (4.3), the result is $Q^p - Q^n > 0$ and is numerically precisely that which is found in ordinary BCS theory near $T_c$. This is no surprise; the same uniform $\mathcal{D}$ gives the BCS expression when used in the $T_c$ equation, Eq. (3.14). If we spread out the delta functions, as would occur for a more realistic magnon propagator, the sign of $Q^p - Q^n$ will not change, the superfluid density will be positive with a value intermediate between zero and the BCS number.

As an illustration, we consider a magnon propagator describing a resonance, as before, but one which is spatially uniform:

$$\mathcal{D}(\mathbf{q},\nu) = -\delta_\mathbf{q} \frac{\Gamma}{(i\nu)^2 - \Gamma^2}. \tag{4.5}$$

When this is used to evaluate Eq. (4.3), we find a result which may be smaller than the BCS value, and, we emphasize, *of the same sign*:

$$\frac{Q_{odd}}{Q_{BCS}} = 19.4 \frac{T_c}{\Gamma}. \tag{4.6}$$

This result has been obtained as the leading term in an expansion in $T_c/\Gamma$ which has been assumed small. Therefore, there is a Meissner effect for the odd frequency composite condensate with a reduction of the superfluid density from the BCS value by a factor of order of the ratio of the transition temperature to the resonance width. We have also checked the superfluid density for the case that the magnon propagator $\mathcal{D}$ sharply peaked in momentum space at either $\mathbf{q} = 0$ or $\mathbf{q} \sim (\pi/a, \pi/a, \pi/a)$. In each case it is again positive and it is somewhat smaller for the second case.

The calculation of other response functions is similar to that for the Meissner effect and we give an example in Section VI.

## V. JOSEPHSON EFFECT BETWEEN ODD AND BCS SUPERCONDUCTORS

Here we briefly consider the Josephson effect between odd and even (BCS) superconductors. To be specific, consider an S-I-S junction between an odd-frequency triplet superconductor on the left and an even-frequency triplet superconductor on the right. By applying the standard tunneling hamiltonian to this system, we find that the tunneling current has the form:

$$J \sim \sum_{\mathbf{k},\mathbf{k}p,\omega} |t_{\mathbf{k},\mathbf{k}p}|^2 F_{odd}(\mathbf{k},\omega) F_{even}(\mathbf{k}p,\omega) \tag{5.1}$$

$$+ \sum_{\mathbf{k},\mathbf{k}p,\mathbf{k}'',\mathbf{k}''',\omega} |t_{\mathbf{k},\mathbf{k}'}|^2 |t_{\mathbf{k}'',\mathbf{k}'''}|^2 F_{odd}(\mathbf{k},\omega) F_{odd}(\mathbf{k}'',\omega) F_{even}(\mathbf{k}p,\omega) F_{even}(\mathbf{k}''',\omega),$$

where $t_{\mathbf{k},\mathbf{k}p}$ is the hopping matrix element and we retain only the relevant fourth order term in the tunneling matrix element expansion.[18] For tunneling between either odd-odd or even-even gap superconductors the first term in the above equation is nonzero and dominates the Josephson current. The situation is drastically different in the present case. As long as there is no inelastic scattering in the junction, the leading term in the Josephson current between odd and even superconductor vanishes since the frequency integral in the first term is *exactly* zero.. However, the next term is nonzero and produces a current which corresponds to the tunneling of "pairs of pairs" with charge $4e$. The very important implication of this result is for flux quantization. If we consider a combined ring, made out of even and odd superconductors, then the flux quantization inside this ring will be in units of

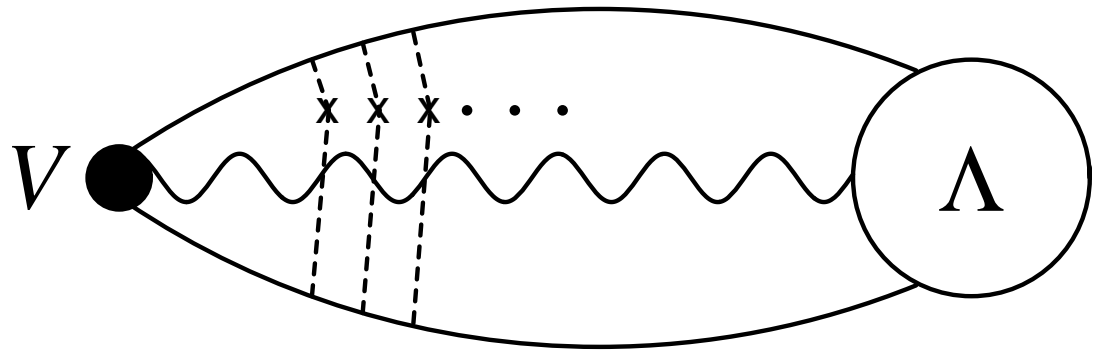

FIG. 9. Linearized gap equation in the presence of impurities. Solid lines are impurity averaged Green's functions. Dotted lines are electron-impurity interactions.

The $T_c$ equation which replaces Eq. (3.14) is

$$1 = VT_c^2 \sum_{\mathbf{k},\mathbf{k}'} \sum_{\omega,\omega'} \mathcal{G}(\mathbf{k},\omega)\mathcal{G}(\mathbf{k}',\omega')\mathcal{D}(\mathbf{k}+\mathbf{k}',\omega+\omega')P(\mathbf{k}+\mathbf{k}',\omega+\omega'), \tag{7.1}$$

where $P(\mathbf{k}+\mathbf{k}',\omega+\omega')$ is the impurity vertex ladder seen in Fig. 9. It is given by

$$P(\mathbf{k}+\mathbf{k}',\omega+\omega') = \tau[D\ (\mathbf{k}+\mathbf{k}')^2 + (\omega-\omega')s] \tag{7.2}$$

Here, $1/\tau$ is the impurity scattering rate, $D$ is the diffusion constant $v_F^2\tau/3$ and $s = [sgn(\omega) - sgn(\omega')]/2$. This result for the vertex is appropriate in the hydrodynamic regime $|\mathbf{k}+\mathbf{k}'| < 1/v_F\tau$ and $\omega - \omega' < 1/2\tau$. Otherwise, $P = 1$.

When Eq. (7.1) is evaluated with either of the magnon propagators mentioned in Section IIIC, i.e. the momentum-independent one or the spatially uniform one, finds a $T_c$ which is unchanged from the clean case up to corrections of order $(1/E_F\tau)^2$.

## VIII. ACKNOWLEDGEMENTS

Part of the work reported here was performed under the auspices of the Program on Strongly Correlated Electron Systems at the Los Alamos National Laboratory. The hospitality of the Aspen Center for Physics, where part of the work was carried out is acknowledged by EA and AVB. AVB was supported in part by a J.R. Oppenheimer Fellowship at Los Alamos. The support of NSF grants DMR92-21907 (EA), DMR92-25027 (DJS) and DMR92-22682 (JRS) is also acknowledged. We are grateful to the following for helpful discussions and useful criticism: P. Coleman, A. Dugaev, G.M. Eliashberg, L.P. Gor'kov, and P. Wölfle.